\documentclass[conference]{IEEEtran}
\usepackage{hyperref}
\usepackage{booktabs}

\usepackage{amsmath,amssymb,amsfonts}
\usepackage{amsmath,amssymb,amsfonts}
\usepackage{mathtools}
\usepackage{graphicx}
\usepackage{xpatch}
\usepackage{dblfloatfix}    

\usepackage{adjustbox,lipsum}

\usepackage[flushleft]{threeparttable}

\usepackage{subcaption}
\usepackage{placeins}

\usepackage[font=footnotesize]{subcaption}

\usepackage[utf8]{inputenc}

\usepackage{sansmath}
\usepackage[defaultmathsizes, italic]{mathastext}

\usepackage{ifdraft}
\usepackage{etoolbox}
\usepackage{xspace}

\usepackage[per-mode=symbol,group-separator={,},detect-all, list-final-separator={, and }]{siunitx}

\usepackage{algorithm}
\usepackage{algorithmicx}
\usepackage{algpseudocode}

\usepackage{tikz}
\usetikzlibrary{positioning}
\usetikzlibrary{backgrounds,scopes}
\usetikzlibrary{fit}
\usetikzlibrary{shapes}
\usetikzlibrary{matrix}
\usetikzlibrary{math}
\usetikzlibrary{calc}
\usetikzlibrary{arrows.meta}
\usetikzlibrary{decorations.pathreplacing}
\usetikzlibrary{calligraphy}
\usetikzlibrary{external}
\usetikzlibrary{patterns}

\usepackage{tikzpagenodes}
\usepackage[siunitx]{circuitikz}
\usepackage{pgfplots}
\pgfplotsset{compat=1.18} 
\usepackage{circledsteps}
\newcommand\myCircled[2][]{\ifmmode%
\Circled[fill color=black,inner color=white,#1]{\footnotesize\mathsf{#2}}%
\else%
\Circled[fill color=black,inner color=white,#1]{\footnotesize\sffamily#2}%
\fi%
}

\tikzset{
    font={\fontsize{8pt}{9}\selectfont},
    arrow/.style={-latex}
    }

\usepackage[noabbrev,capitalise]{cleveref}
\Crefname{subsection}{Section}{Sections}
\Crefname{subsubsection}{Section}{Sections}
\Crefname{paragraph}{Section}{Sections}
\Crefname{figure}{Fig.}{Fig.}
\Crefname{table}{Tab.}{Tab.}

\usepackage{microtype}
\usepackage[normalem]{ulem}
\usepackage{tabularx}
\usepackage{booktabs}
\usepackage{multirow}
\usepackage{makecell}
\usepackage{textcomp}

\usepackage{enumitem}
\usepackage{balance}

\usepackage{graphicx}
\usepackage{booktabs}
\usepackage{multirow}
\usepackage{multicol}
\usepackage{array}
\usepackage{tabularx}
\usepackage{bm}
\usepackage{soul}
\usepackage{tikz}
\usepackage{amssymb}
\usepackage{amsmath}
\usepackage{xfrac}
\usepackage[bottom]{footmisc}
\usepackage{fixltx2e}
\usepackage{amsthm}
\usepackage{algorithm}
\usepackage{amsmath, amssymb}
\usepackage{algorithmicx}
\usepackage{framed}

\theoremstyle{definition}

\theoremstyle{remark}

\definecolor{dgreen}{RGB}{0, 139, 0}
\usepackage{enumitem}


\let\oldbibliography\thebibliography
\renewcommand{\thebibliography}[1]{\oldbibliography{#1}
\setlength{\itemsep}{-1pt}} 

\usepackage{multirow}
\usepackage{colortbl} 
\usepackage[table]{xcolor}
\usepackage{siunitx}
\usepackage{graphicx}
\usepackage{subcaption}
\usepackage{svg}
\usepackage[a4paper, total={184mm,239mm}]{geometry}

\def\BibTeX{{\rm B\kern-.05em{\sc i\kern-.025em b}\kern-.08em
    T\kern-.1667em\lower.7ex\hbox{E}\kern-.125emX}}
\begin{document}

    \title{
        ASTER: Attention-based Spiking Transformer Engine for Event-driven Reasoning
    }
    \pagestyle{plain}
    \author{
    \IEEEauthorblockN{Tamoghno Das, Khanh Phan Vu, Hanning Chen, Hyunwoo Oh, Mohsen Imani}
    \IEEEauthorblockA{\textit{\{tamoghnd, vukp1, hanningc, hyunwooo, m.imani\}@uci.edu}}
    \IEEEauthorblockA{\textit{Department of Computer Science, University of California, Irvine, USA}}
}
    
\maketitle

\begin{abstract}
    The integration of spiking neural networks (SNNs) with transformer-based architectures has opened new opportunities for bio-inspired low-power, event-driven visual reasoning on edge devices. However, the high temporal resolution and binary nature of spike-driven computation introduce architectural mismatches with conventional digital hardware (CPU/GPU). Prior neuromorphic and Processing-in-Memory (PIM) accelerators struggle with high sparsity and complex operations prevalent in such models. To address these challenges, we propose a memory-centric hardware accelerator tailored for spiking transformers, optimized for deployment in real-time event-driven frameworks such as classification with both static and event-based input frames. Our design leverages a hybrid analog-digital PIM architecture with input sparsity optimizations, and a custom-designed dataflow to minimize memory access overhead and maximize data reuse under spatiotemporal sparsity, for compute and memory-efficient end-to-end execution of spiking transformers. We subsequently propose inference-time software optimizations for layer skipping, and timestep reduction, leveraging Bayesian Optimization with surrogate modeling to perform robust, efficient co-exploration of the joint algorithmic-microarchitectural design spaces under tight computational budgets. Evaluated on both image(ImageNet) and event-based (CIFAR-10 DVS, DVSGesture) classification, the accelerator achieves up to ~467x and ~1.86x energy reduction compared to edge GPU (Jetson Orin Nano) and previous PIM accelerators for spiking transformers, while maintaining competitive task accuracy on ImageNet dataset. This work enables a new class of intelligent ubiquitous edge AI, built using spiking transformer acceleration for low-power, real-time visual processing at the extreme edge. 
\end{abstract}

\section{Introduction}


Spiking Transformers combine the temporal sparsity and low-precision encoding of Spiking Neural Networks (SNNs)~\cite{maass1997networks, roy2019towards} with the long-context modeling of self-attention~\cite{vaswani2017attention}. Unlike conventional Transformers requiring high-bit MACs, they process sparse binary spike trains, replacing most matrix operations with masked additions or bitwise logic—making them attractive for energy- and latency-constrained edge AI.


Dynamic Vision Sensors (DVS) generate asynchronous, event-driven streams unsuitable for frame-based Transformers. Spiking Transformers, however, naturally handle such inputs via integrate-and-fire (LIF) dynamics, propagating activations only when thresholds are crossed. Recent models like Spikformer~\cite{zhou2022spikformer}, SDT~\cite{yao2023spike}, and SpikingResFormer~\cite{shi2024spikingresformer} achieve high accuracy on classification and detection with far lower energy than Vision Transformers (ViTs). Notably, SDT removes non-spiking residuals, enabling full integer quantization and spike-only inference.

\begin{figure}[t!]
    \centering
    \includegraphics[width=0.7\linewidth]{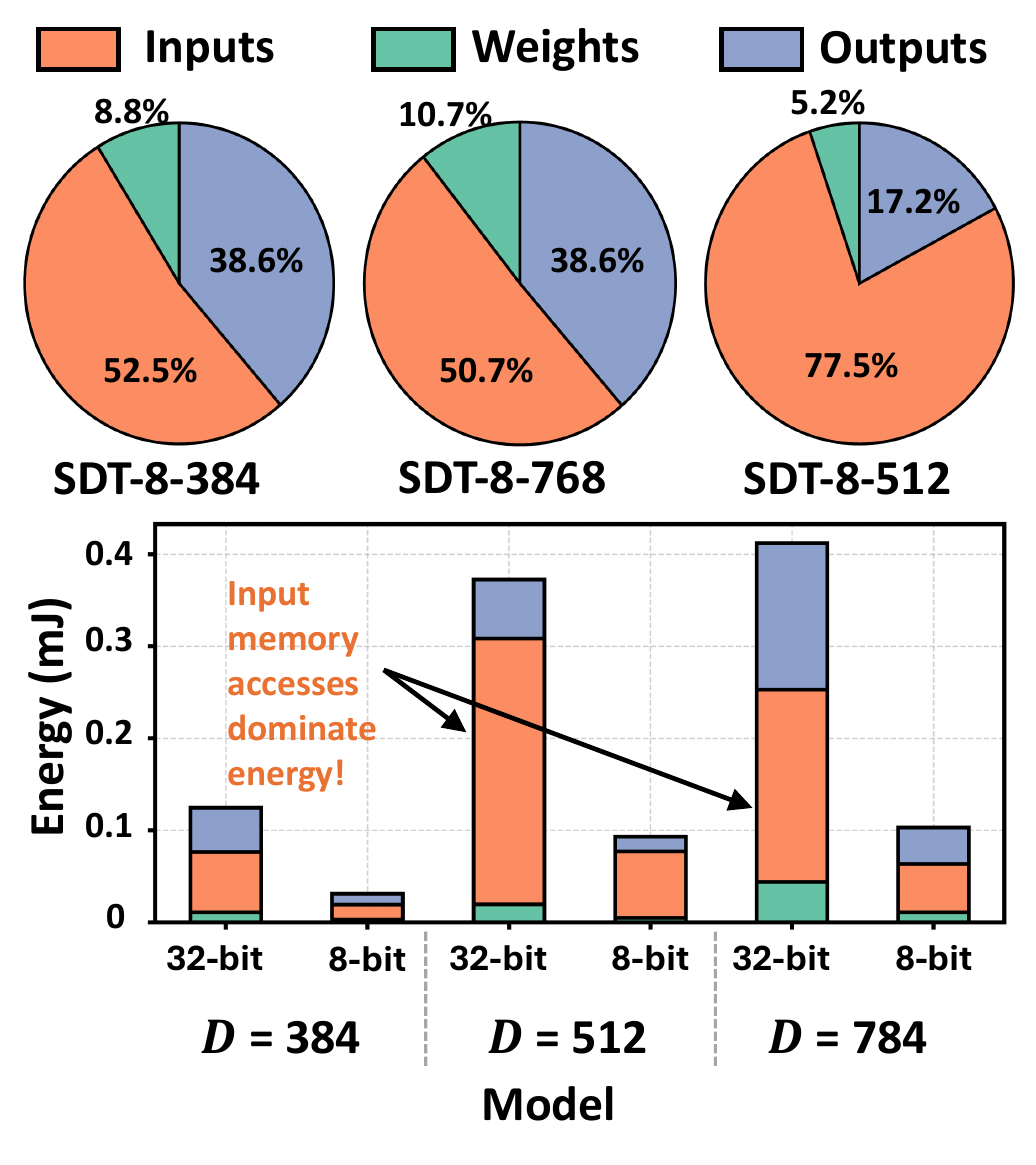}
    \vspace{-3mm}
    \caption{Memory Access Distribution for Patch Embedding in Spiking Transformer on ImageNet}
    \label{fig:memory_accesses_breakdown}
\end{figure}

\vspace{-1mm}


However, conventional CPUs and GPUs struggle with the fine-grained sparsity of spiking models, incurring high memory traffic and lacking asynchronous control. Processing-in-Memory (PIM), particularly resistive crossbar (RRAM)-based designs, alleviates this by supporting in-situ accumulation and reduced data movement—well-suited to spike-driven workloads.


Mapping Spiking Transformers to PIM remains challenging: (i) retaining temporal state across LIF neurons without frequent memory accesses; (ii) implementing bitwise attention (e.g., SDSA~\cite{yao2023spike}) through local masking and accumulation; and (iii) aligning irregular event-driven inputs with spatially tiled PIM arrays.

To address these gaps, we propose \textbf{ASTER}, a hardware–software co-designed PIM accelerator tailored for Spiking Transformers. On the software side, we benchmark spike activity and apply unified pruning across layers and timesteps. On the hardware side, ASTER integrates sparse-aware compute-in-memory (CIM) crossbars for spike accumulation and self-attention, along with time-multiplexed membrane buffers for LIF dynamics. The dataflow supports both spike-stationary and weight-stationary modes to maximize reuse across space and time. ASTER is further designed to interface with neuro-symbolic reasoning backends, such as Hyperdimensional Computing (HDC), enabling symbolic post-processing over spike-encoded representations.

\textbf{Contributions:}
\begin{enumerate}
    \item We present the first end-to-end compute-in-memory accelerator for Spiking Transformers, supporting fully event-driven execution of attention and MLP blocks in analog PIM arrays.
    \item We propose temporal reuse via membrane state persistence and weight sharing across timesteps to minimize redundant accesses.
    \item We design a spike-aware self-attention datapath using bitwise masking and column-wise reduction, with support for time-averaged classification in memory.
    \item We introduce task-aware layer skipping and confidence-based dynamic timestep reduction to exploit runtime sparsity without sacrificing accuracy.
\end{enumerate}

ASTER achieves a 1.86$\times$ energy improvement over prior PIM baselines~\cite{song2025xpikeformer}, and up to 467$\times$ over GPU baselines on event-driven visual classification tasks.
    
    \section{RELATED WORK}
\label{sec:related}

\subsection{Spiking Transformers}

Spiking Neural Networks (SNNs) offer a low-power alternative to ANNs by using temporally sparse, binary spike-based communication~\cite{maass1997networks,roy2019towards}. Spiking Transformers integrate self-attention~\cite{vaswani2017attention} into SNNs to capture long-range temporal dependencies. Spikformer~\cite{zhou2022spikformer} introduced a spiking self-attention module using multiplications on sparse spike matrices, while the Spike-Driven Transformer (SDT)~\cite{yao2023spike} improved efficiency by replacing dot-products with masked additions and using membrane potential shortcuts for residuals. These models achieve competitive results on tasks like classification and gesture recognition but are usually tested on GPUs, which underutilize sparsity. Event-based vision systems such as Dynamic Vision Sensors (DVS) generate sparse temporal spike streams for applications like gesture recognition~\cite{amir2017low} and visual question answering~\cite{chen2024vqa}. Current Spiking Transformers, however, rely on ANN-to-SNN conversions with large time-steps~\cite{ann_snn_conversion} or hybrid training that limits full exploitation of neuromorphic hardware’s address‑based compute~\cite{aydin2024hybrid, zhong2025hynita}.
ASTER adopts SDT~\cite{yao2023spike} as a spike-native, integer-quantized frontend due to its fully event-driven architecture and compatibility with hardware execution. Unlike prior work proposing new attention mechanisms, ASTER focuses on co-designing a hybrid analog–digital Processing-in-Memory (PIM) backend that natively executes spike-driven attention and MLP layers under temporal sparsity and binary encoding.

\subsection{Event-driven Hardware Accelerators}
Mainstream neuromorphic platforms such as Intel’s Loihi~\cite{davies2018loihi} and IBM’s TrueNorth~\cite{akopyan2015truenorth} provide highly efficient event‑driven acceleration for generic SNN layers, but offer no hardware‑native support for transformer‑style self‑attention primitives~\cite{yao2023spike}.  In the recent past, some works have tried to address this gap.  
Xu et al.~\cite{xu2024spiking} propose a 3D‑stacked logic‑on‑logic and memory‑on‑logic accelerator that maximizes weight reuse and minimizes interconnect energy for spiking self‑attention, demonstrating significant energy and latency improvements over equivalent 2D CMOS designs. Chen and Chang~\cite{chen2025tickbatching} introduce a low‑power ASIC employing “tick‑batching,” in which all simulation time‑steps execute in parallel and residual addition is replaced by a bitwise IAND.    
Chen et al.~\cite{chen2024vesta} present VESTA, a unified processing‑element fabric that flexibly supports convolution, MLP, and spiking self‑attention by time‑multiplexing PEs in an event‑driven pipeline, yielding high utilization across diverse transformer variants.
None of these accelerators, however, provides direct hardware support for the sparse mask‑add and multi‑head accumulation operations central to spiking self‑attention, motivating our PIM‑centric architecture.

    \section{ASTER Software Design}

\subsection{Sparsity-Aware Model Optimizations}
\begin{figure}[h!]
    \centering
    \includegraphics[width=0.8\linewidth]{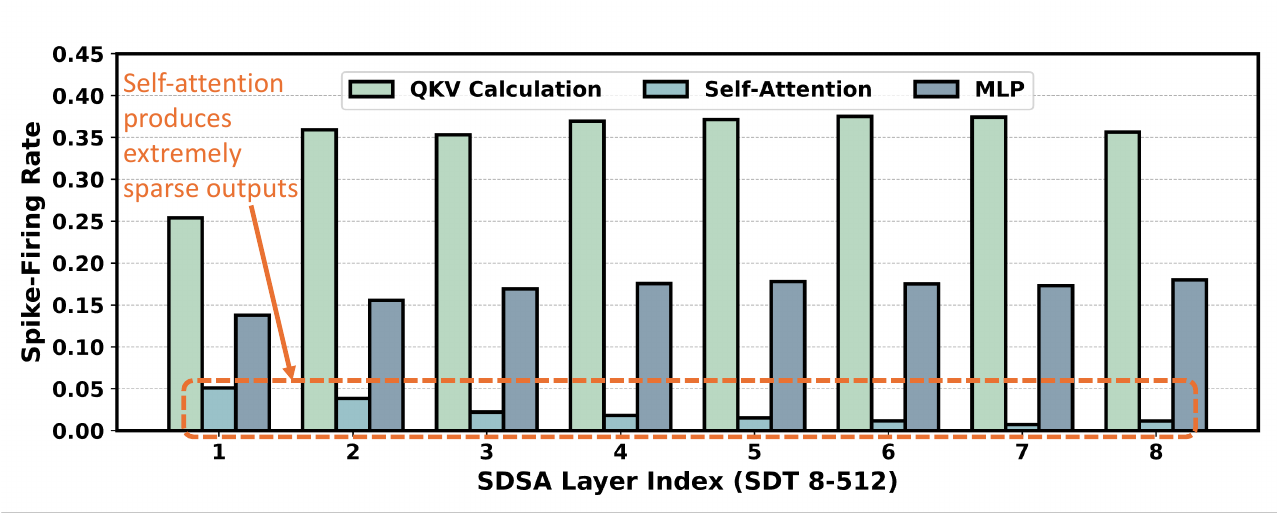}
    \caption{Layerwise Spike Firing Rates for SDT-8-512, illustrating extremely high sparsity for outputs of spiking self-attention}
    \label{fig:layer_firing_rate}
\end{figure}

\begin{figure}[h!]
    \centering
    \includegraphics[width=\linewidth]{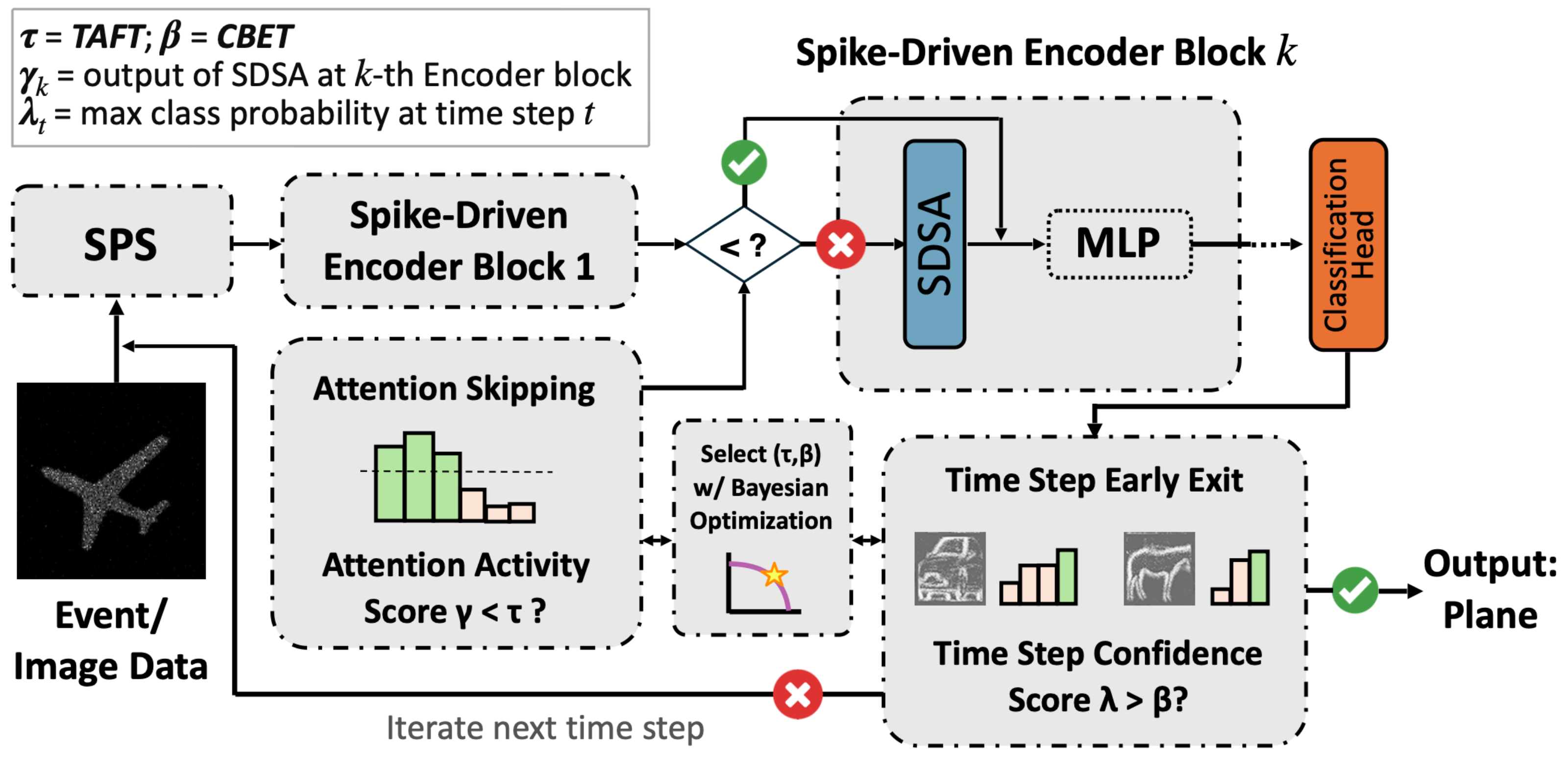}
    \caption{Software-side framework with layer skipping by attention activity and timestep reduction by early exit.}
    \label{fig:sw_framework}
\end{figure}

LIF activations increase sparsity in Spiking Transformers~\cite{yao2023spike}, which can undermine efficiency if unmanaged. As shown in Fig.~\ref{fig:layer_firing_rate}, firing rates in deeper self-attention layers decline, and when rates fall below $5\%$~\cite{zhang2024you}, their contribution to classification becomes negligible. Moreover, not all inputs require the full temporal sequence: simple samples reach high-confidence predictions with fewer timesteps, while complex ones need the entire window. Guided by these observations, we introduce three software-level optimizations that exploit inherent sparsity to improve spike-driven Transformer efficiency (Fig.~\ref{fig:sw_framework}).

\textbf{Task-Aware Adaptive Layer Skipping:}  We introduce a method to selectively bypass attention blocks characterized by consistently low spike activity. During a preliminary validation phase, spike activity patterns are profiled across layers (Fig. \ref{fig:layer_firing_rate}), identifying those below a defined \textbf{\textit{Task-Aware Firing Threshold (TAFT)}}. Identified blocks perform identity transformations during inference, thus eliminating unnecessary QKV computations. Any combination of layers can be skipped based on the specific sparsity profiles identified. Layer skipping decisions are made once at model initialization to maintain low runtime overhead and ensure compatibility with existing pre-trained architectures.

\begin{figure}
    \centering
    \includegraphics[width=0.85\linewidth]{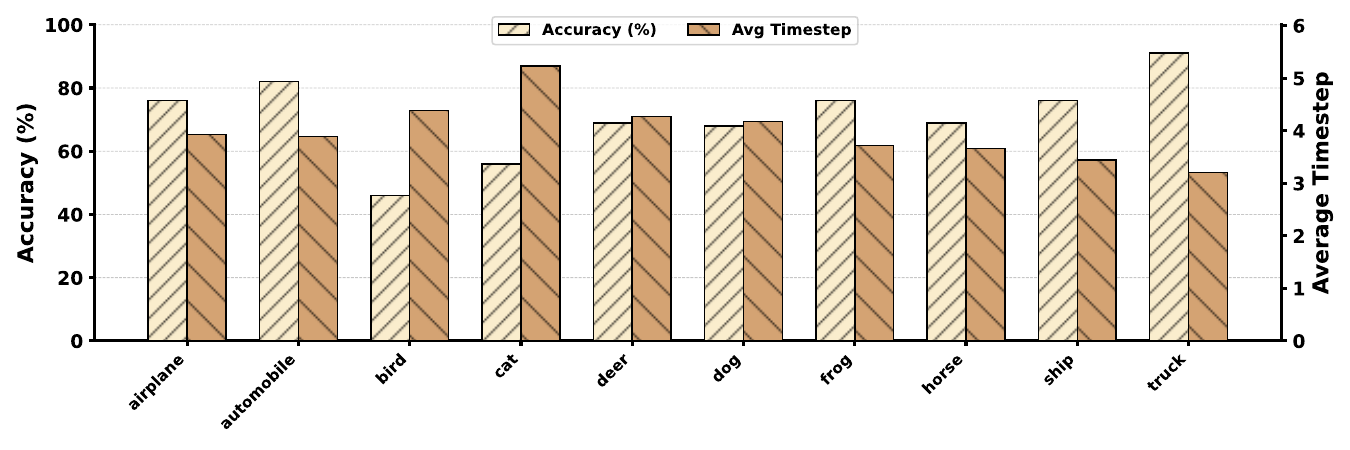}
    \caption{Per-class Average Timestep vs Accuracy for 2-256-t16 SDT on CIFAR10-DVS with CBET=0.99.}
    \label{fig:early_exit_class_stats}
\end{figure}

\textbf{Confidence-Based Dynamic Timestep Reduction:} Taking inspiration from \cite{li2023seenn}, this optimization dynamically controls the number of timesteps based on prediction confidence using the maximum softmax probability or normalized entropy metrics based on a randomly chosen subset of input data. Accumulated logits over multiple timesteps are continuously evaluated, and the inference process terminates early once prediction confidence surpasses a defined \textbf{\textit{Confidence-Based Exit Threshold (CBET)}}. Typically, simpler inputs achieve high confidence rapidly, reducing computational demands. As shown in Fig. \ref{fig:early_exit_class_stats}, we provide a per-class timestep analysis that enables application-specific threshold tuning based on the statistical distribution of each input class to balance computational efficiency against model accuracy.

\textbf{Combining Layer Skipping and Timestep Reduction:} The two features can be deployed independently or in combination to achieve synergistic power savings. The modular design ensures broad applicability across architectures without requiring architectural modifications or retraining, with configurable thresholds allowing fine-tuning of the accuracy-efficiency trade-off for specific deployment scenarios.

\subsection{Optimizing selection of TAFT and CBET}

\begin{figure}
    \centering
    \includegraphics[width=0.9\linewidth]{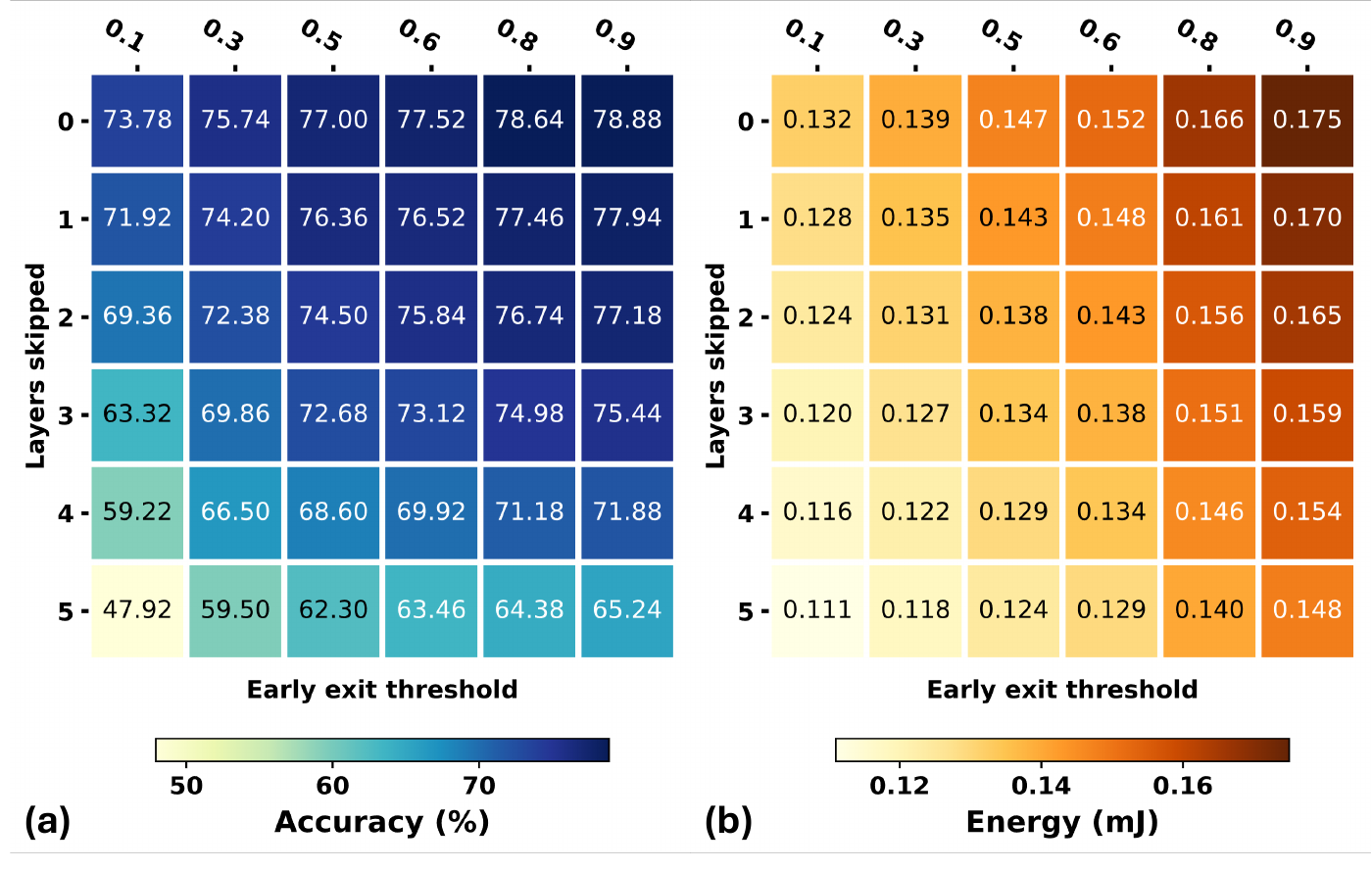}
    \caption{Combined effect of layer skipping and early exit on a) Accuracy, and b) Energy.}
    \label{fig:sdt_accuracy_energy_heatmap}
\end{figure}

ASTER utilizes two dynamic pruning mechanisms for efficiency: SDSA layer skipping, controlled by the Task-Aware Firing Threshold (TAFT, denoted $\tau$), and temporal early exit, governed by the Confidence-Based Exit Threshold (CBET, denoted $\beta$). The interaction between these thresholds determines the energy–accuracy trade-off (shown later in Fig.~\ref{fig:sdt_accuracy_energy_heatmap}).

However, tuning $(\tau, \beta)$ manually is labor-intensive, dataset-specific, and fails to fully capture their joint interaction. Instead, to generalize threshold selection and tuning, we formulate a bi-objective optimization problem that seeks Pareto-optimal pairs $(\tau, \beta)$ maximizing both task accuracy and energy efficiency. For each candidate configuration $\theta = (\tau, \beta)$, we define a scalar \textbf{objective function}:
\begin{equation}
g(\theta) = \alpha \cdot \mathrm{Acc}(\theta) - (1 - \alpha) \cdot \mathrm{E}_{\text{norm}}(\theta)
\label{eq:scalar_objective}
\end{equation}

Here, $\mathrm{Acc}(\theta)$ and $\mathrm{E}_{\text{norm}}(\theta)$ are the normalized accuracy and energy metrics under threshold setting $\theta$, and $\alpha \in [0,1]$ governs the trade-off. We apply Bayesian Optimization \cite{bayesianoptimization} over the space $\Theta = [\tau_{\min}, \tau_{\max}] \times [\beta_{\min}, \beta_{\max}]$ using Expected Improvement as the acquisition function, with a Gaussian Process surrogate model $g(\theta)$. Final configurations lie on Pareto frontier $\mathcal{P} \subset \Theta$, providing deployment-ready thresholds tailored to task constraints and priorities (detailed in Algorithm ~\ref{alg:bayesianalgo}).

\section{ASTER Architecture Design}

\subsection{Processing-in-Memory Architecture}

\begin{figure*}[h!]
    \centering
    \includegraphics[width=0.8\linewidth]{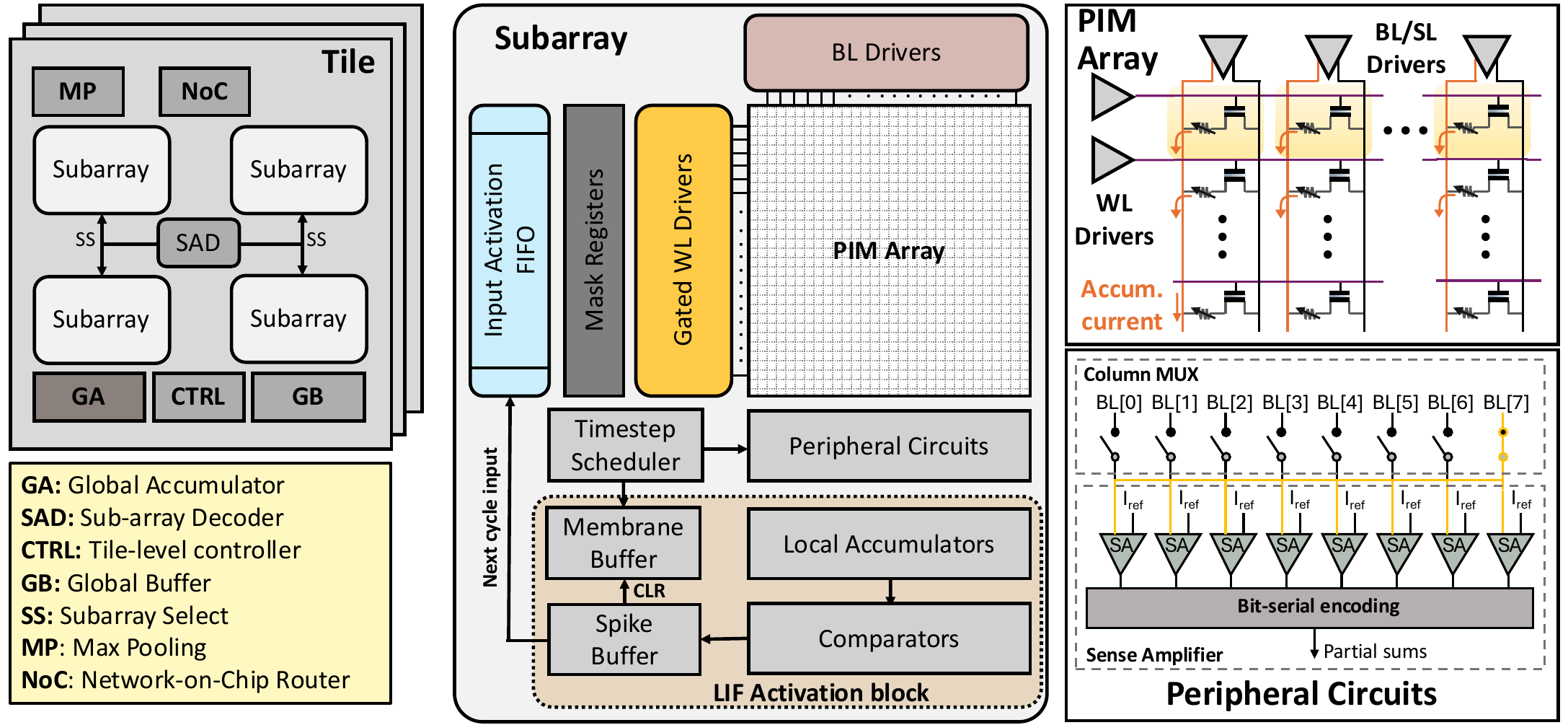}
    \caption{Overall hardware architecture of ASTER, showing hierarchy at chip, tile, and subarray levels.}
    \label{fig:main_hw_arch}
\end{figure*}

To harness the event-driven nature of spiking computation, the ASTER hardware architecture (Fig. \ref{fig:main_hw_arch}) adopts a hierarchical organization. The chip consists of multiple tiles, each coordinating computation across several RRAM-based crossbars known as subarrays. Subarray selection is orchestrated through Subarray Select (SS) signals from Subarray decoders (SADs), enabling coarse-grained power gating by activating only the selected subarray's high-voltage wordline rails.

Within each subarray, input vectors are broadcast for in-memory analog matrix-vector multiplications, which leverage current accumulation and sparse spike encoding to reduce peripheral switching. The results are thresholded to execute local LIF dynamics and membrane updates with low latency. In parallel, the tile-level controller manages global data orchestration and aggregates partial results via a global accumulator. This architecture confines high-bandwidth analog computation to the subarray level while sharing low-duty-cycle digital logic at the tile level, creating an energy-efficient pipeline that fully leverages input sparsity.

\subsection{Programmable Selective WL Activation to exploit input sparsity}



Spiking Transformers use Leaky-Integrate-and-Fire (LIF) activations that produce highly sparse spike streams. Naively mapping these to a PIM array toggles every word-line (WL) each cycle, incurring disproportionate energy costs. As shown in Fig. \ref{fig:memory_accesses_breakdown}, memory access energy is dominated by inputs. Prior schemes \cite{datta2022ace, perera2025low, putra2021spikedyn} use coarse bank gating \cite{datta2022ace} or require large AER buffers \cite{perera2025low}.

\begin{figure}[ht!]
    \centering
    

    
    \includegraphics[width=\linewidth]{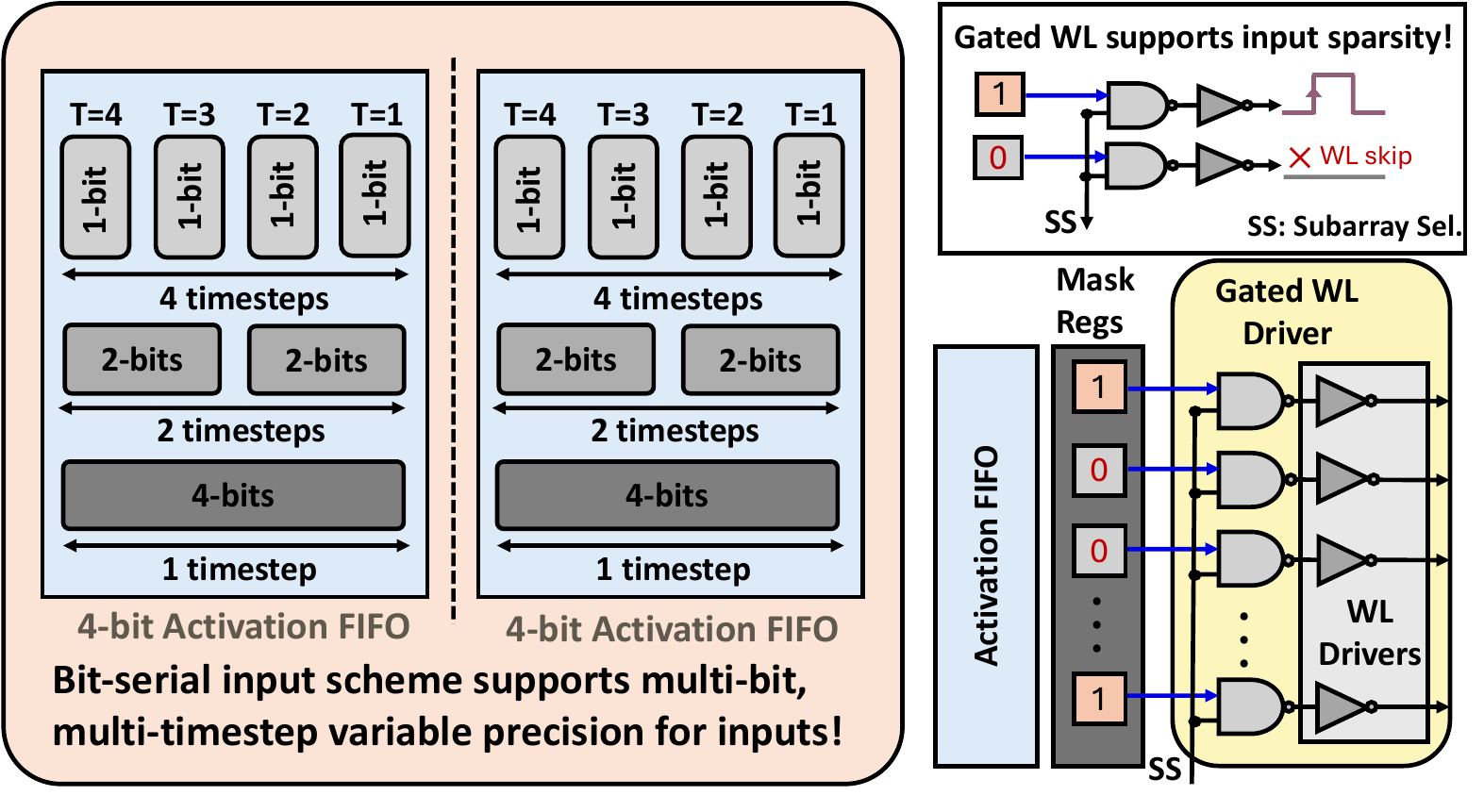}
    

    \caption{Hybrid input support in ASTER PIM supporting Multi-bit mapping via bit-serial FIFOs and masking, with gated WL execution for sparse row skipping.}
    \label{fig:hybrid_input_support}
\end{figure}

In contrast, we introduce a fine-grained, programmable WL-driver that interprets the input vector as a per-row mask (Fig. \ref{fig:hybrid_input_support}). The driver asserts only WLs corresponding to non-zero activation bits, eliminating redundant MAC operations and reducing energy and latency. This zero-skipping logic is implemented with minimal area overhead and is agnostic to the memory technology. 

Embedding this logic into a bit-serial PIM datapath supports both spiking and conventional DNNs. Activations are streamed one bit-plane at a time, and a WL fires only if the corresponding bit is '1'. Energy and latency therefore scale with the input's Hamming weight rather than its bit-width. This allows for single-cycle spike decisions in SNNs and adaptive precision in multi-bit CNNs/MLPs using the same hardware at no extra cost.

\subsection{Custom Dataflow \& SSA Module}

\begin{figure}
    \centering
    \includegraphics[width=0.9\linewidth]{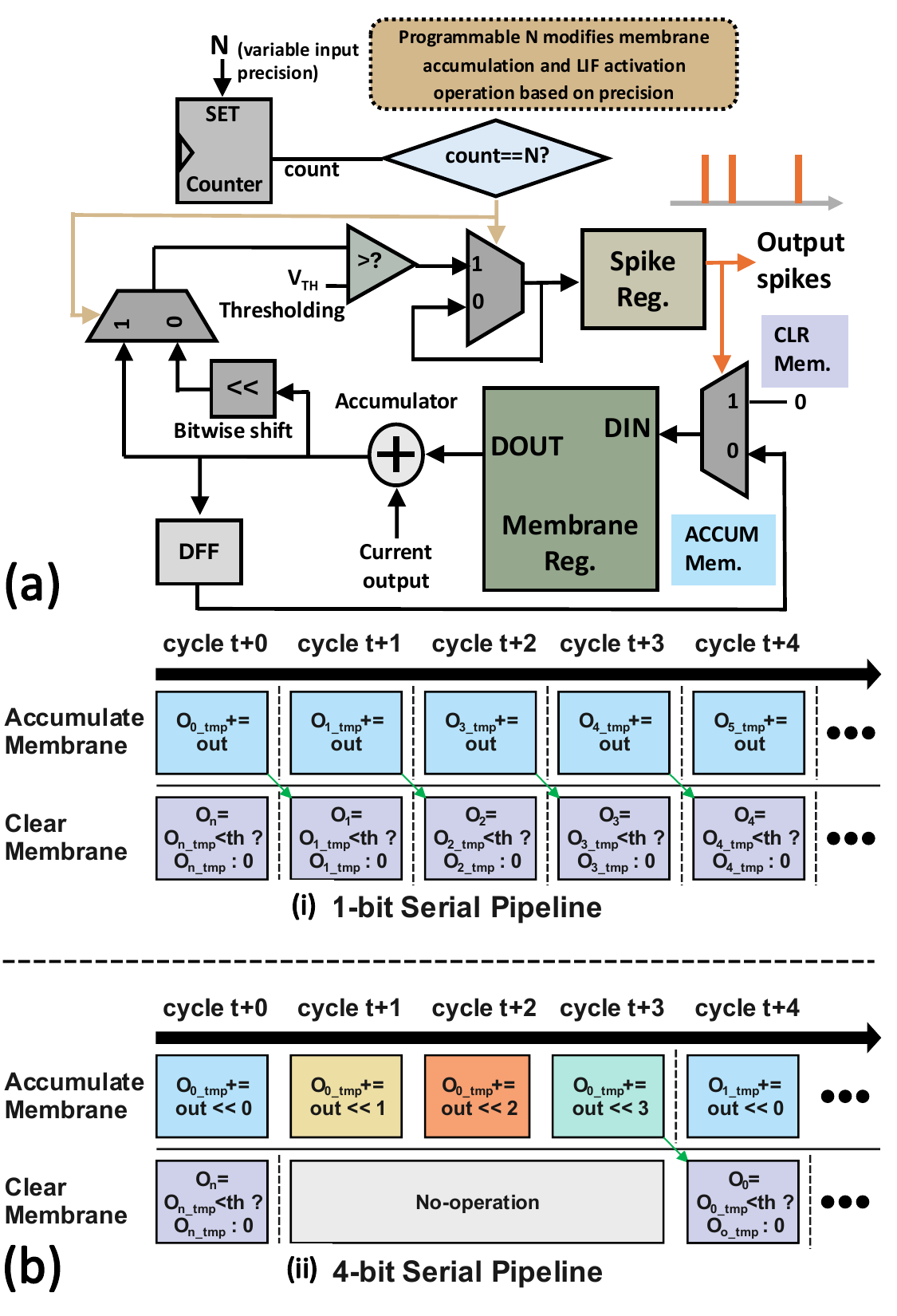}
    \caption{(a) LIF Activation and Membrane Potential Accumulation block, which supports variable bit-precision using bit-serial computing. (b) Pipeline showing difference between 1-bit and 4-bit input activations, on membrane register updates.}
    \label{fig:lif_accum}
\end{figure}

We propose a dataflow tailored to the reuse patterns in Spiking Transformers. The core Spike-Driven Self-Attention (SDSA) module computes attention maps ($Q \odot K$) using a binary mask-and-add operation, realized with bitwise ANDs and conditional accumulations. These operations map naturally to RRAM arrays using masked row activations, ensuring sublinear energy scaling with input spike sparsity. To maximize efficiency, we adopt a hybrid mapping policy:
\begin{itemize}
    \item \textbf{Weight-stationary mode} for patch embedding and MLP layers, enabling persistent caching of weights in PIM arrays across timesteps.
    \item \textbf{Spike-stationary mode} during SDSA execution to exploit temporal sparsity and avoid fetching inactive spike vectors.
\end{itemize}
This dual-mode mapping significantly improves utilization and minimizes dynamic energy. Temporal reuse is further enabled through the following mechanisms:

\begin{enumerate}
    \item \textbf{Weight Sharing Across Timesteps:} All transformer block weights are reused across timesteps and maintained in local subarray buffers to amortize memory fetch costs.
    \item \textbf{Locality in Patch Embedding:} Intermediate outputs from Conv+Spike layers are cached in local SRAM to facilitate intra-timestep and inter-patch reuse.
    \item \textbf{Membrane State Buffering:} As shown in Fig. \ref{fig:lif_accum}, membrane potentials are accumulated in PIM-embedded registers, minimizing data roundtrips for efficient LIF dynamics.
    \item \textbf{SDSA Reuse Patterns:} Attention weights are reused across channels each time step, benefiting from high reuse of the Q and K vectors during in-situ CIM computations.
    \item \textbf{Classification Head Averaging:} Class logits are aggregated from per-timestep outputs using a temporal accumulator in memory, performing the final linear projection in-situ with negligible overhead.
\end{enumerate}

The dataflow also supports spatial tiling of patch tokens and pipelined execution over time steps. Tiling improves memory locality, while temporal pipelining overlaps compute and data movement to exploit regularity in spike timing and weight reuse.

\begin{table*}[]
\centering
\begin{minipage}{0.65\textwidth}
    \centering
    \caption{Performance and Memory Characteristics of Spiking Transformer Modules}
    \resizebox{\linewidth}{!}{
    \begin{tabular}{l
                    S[table-format=4.2] 
                    S[table-format=5.0] 
                    S[table-format=1.0]
                    S[table-format=2.8]
                    S[table-format=2.8]
                    S[table-format=5.2]}
    \toprule
    \textbf{Module} & \textbf{Memory Access (MB)} & \textbf{GFLOPs} & \textbf{Timesteps} & \textbf{Arithmetic Intensity} & \textbf{Latency (ms)} & \textbf{Throughput} \\
    \midrule
    SPS              & 1023.52 & 20.982 & 4 & 82.001 & 1.619 & 617.47 \\
    Spike-Driven Encoder  & 1704.41 & 11.099 & 4 & 26.049 & 19.737 & 50.67 \\
    Linear           & 2.21   & 0.002     & 1 & 0.694  & 0.093 & 10719.83 \\
    \midrule
    \textbf{Total}   & 3643.32 & 32.083 & 4 & 35.224 & 24.308 & 41.14 \\
    \bottomrule
    \end{tabular}}
\end{minipage}\hfill
\begin{minipage}{0.3\textwidth}
    \centering
    \caption{Component-wise Area Breakdown}
    \resizebox{\linewidth}{!}{
    \begin{tabular}{|l|c|c|}
    \hline
    \textbf{Component} & \textbf{Per subarray} & \textbf{Area (mm$^2$)} \\
    \hline
    RRAM subarray & 1 & 0.05 \\
    Mask registers & 128 & 0.02 \\
    WL Gating Control logic & 128 & 0.015 \\
    ADC (8 cols sharing) & 16 & 0.1 \\
    WL Drivers & 128 & 0.03 \\
    Buffers & 64 & 0.02 \\
    \hline
    \end{tabular}}
\end{minipage}
\end{table*}

\FloatBarrier

    \section{Experimental Evaluation}
\label{sec:results}
\subsection{Experimental Setup}

\subsubsection{Software setup}We implemented the proposed sparsity-aware optimizations into the Spike-Driven Transformer (SDT) model~\cite{yao2023spike}. For conditional attention bypass, spike activity was profiled during validation to identify layers consistently below the TAFT, which were then bypassed at inference. For timestep early exit, we modified inference to produce per-timestep outputs, terminating once prediction confidence exceeded the CBET. Both thresholds were configurable, enabling evaluation of individual and combined strategies.

We evaluated models on three benchmark datasets with representative spatiotemporal characteristics:

\begin{itemize}
    \item \textbf{CIFAR10-DVS} \cite{cifar10dvs}: event-stream version of CIFAR10, using a compact 2-layer SDT (dim=256, 8 heads, 16 timesteps, CutOut augmentation). Trained for 200 epochs with AdamW optimizer and cosine scheduling.
    \item \textbf{ImageNet100} \cite{imagenet}: a subset of ImageNet (100 classes), trained with an 8-layer SDT (dim=512, 8 heads, 4 timesteps). Trained for 200 epochs with LAMB optimizer and strong augmentations (RandAugment, Mixup, CutMix).
    \item \textbf{DVS-Gesture} \cite{amir2017low}: gesture recognition dataset with 11 classes, trained using a 2-layer SDT (dim=256, 8 heads, 16 timesteps). Training used LAMB with cosine decay for 200 epochs.
\end{itemize}

\subsubsection{Hardware setup}
We evaluate energy and area of our analog RRAM-based PIM architecture using a hierarchical flow with well-established tools. Crossbar-level analog MACs are modeled in NeuroSim~\cite{neurosim}, including peripheral drivers and sense amplifiers. Tile-level buffer and I/O costs are estimated in CACTI~\cite{cacti}, with scaling for spike-driven access frequency and bit-precision. LIF activation energy is captured via analytical models of capacitor charging/discharging and comparator switching~\cite{datta2022ace}. Digital control and routing overheads are derived from standard cell libraries. All models are calibrated with cycle-accurate profiling to reflect realistic event-driven execution, providing a robust assessment of PIM efficiency for spike-driven Transformer workloads.

\subsection{Evaluation Results}
\subsubsection{Conditional Attention Bypass via Spike Activity Thresholding}

\begin{figure}[]
    \centering
    \includegraphics[width=0.9\linewidth]{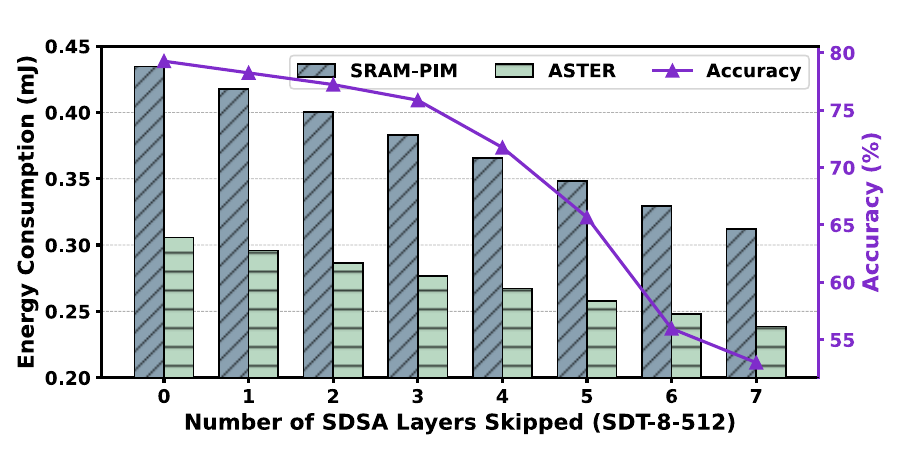}
    \caption{Tradeoff between accuracy and energy consumption due to layer skipping}
    \label{fig:sdt_accuracy_energy_layerskipping}
\end{figure}

Our layer-wise spike analysis revealed pronounced sparsification in self-attention outputs (Fig.~\ref{fig:sdt_accuracy_energy_layerskipping}). While Q, K, and V projections remained relatively stable, post-SDSA firing rates dropped sharply beyond Layer 4, falling below 0.06 (Fig.~\ref{fig:layer_firing_rate}), indicating that attention maps become increasingly sparse due to the Hadamard product of Q and K, leaving the Value outputs ineffective. MLP blocks, however, retained meaningful activations, motivating selective SDSA bypass. Based on this, we selectively skipped deeper attention layers (6–8) using TAFT.

This approach yielded major efficiency gains: ASTER reduced SDSA-related MACs by up to 99.3\%, cutting energy without affecting accuracy while avoiding redundant accumulation and memory access in CIM backends. The bypass decision is fixed at initialization, adding negligible runtime overhead. These results show spike-activity-aware gating effectively leverages intrinsic sparsity, eliminating redundant computation while preserving representational capacity.

\subsubsection{Timestep Early Exit via Confidence Thresholding}

\begin{figure*}[]
    \centering
    \begin{minipage}{0.3\linewidth}
        \includegraphics[width=\linewidth]{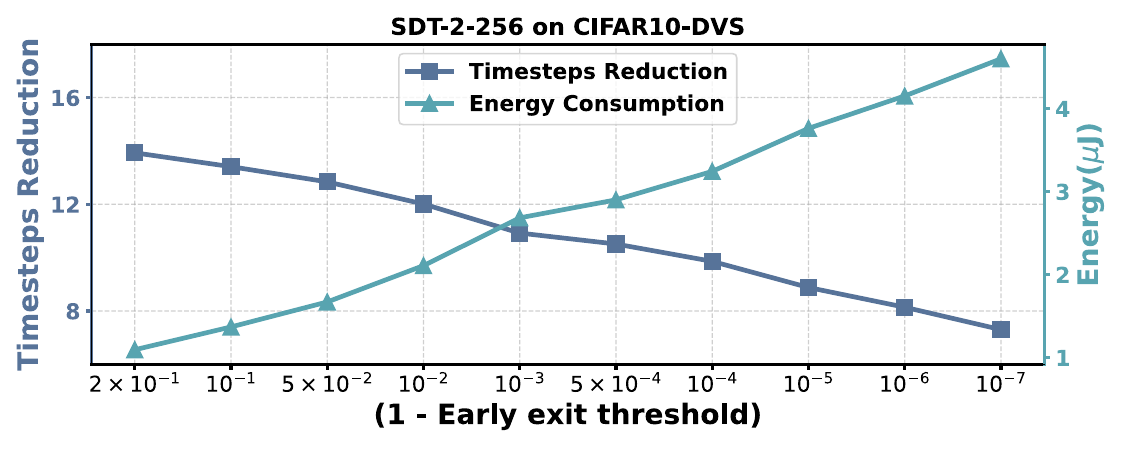}
    \end{minipage}
    \hfill
    \begin{minipage}{0.3\linewidth}
        \includegraphics[width=\linewidth]{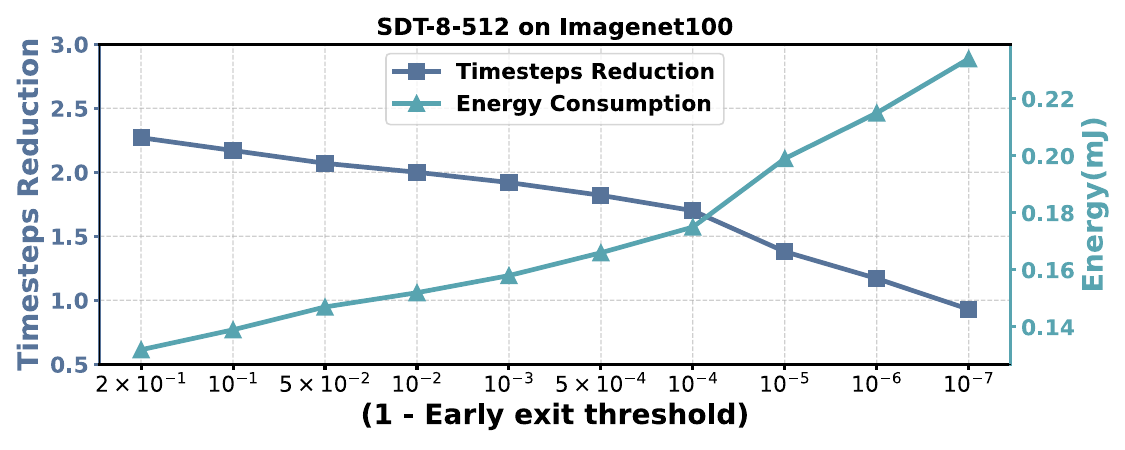}
    \end{minipage}
    \hfill
    \begin{minipage}{0.3\linewidth}
        \includegraphics[width=\linewidth]{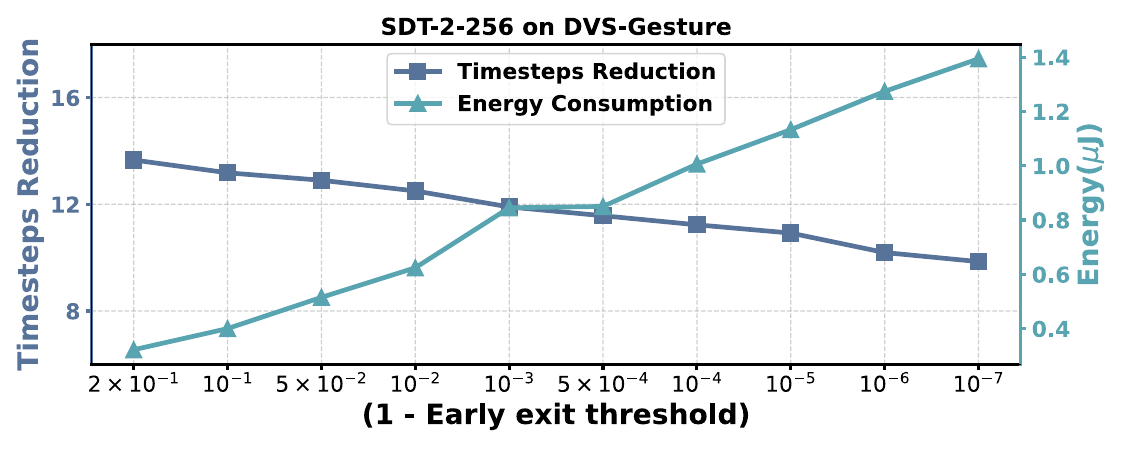}
    \end{minipage}
    \caption{Tradeoff of Early-Exit timestep reduction vs Energy across different datasets.}
    \label{fig:timestep_energy_combined}
\vspace{-3mm}
\end{figure*}

\begin{figure*}[]

    \centering
    \begin{minipage}{0.3\linewidth}
        \includegraphics[width=\linewidth]{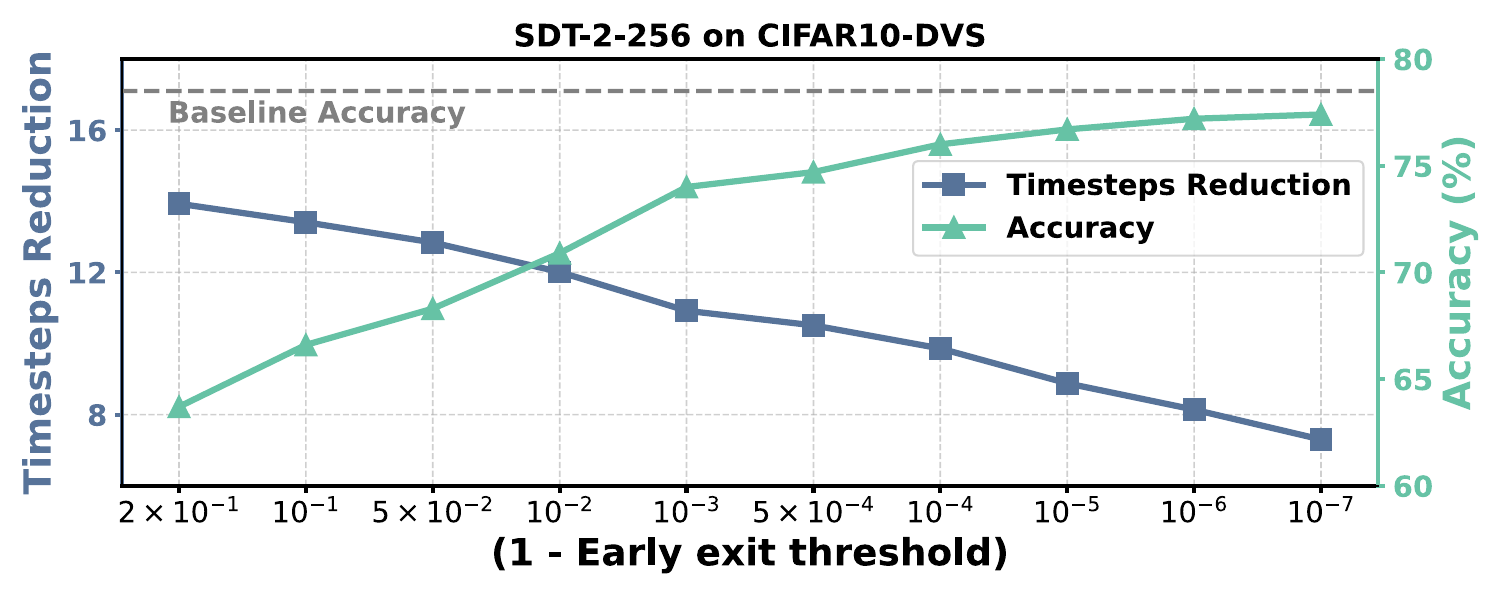}
    \end{minipage}
    \hfill
    \begin{minipage}{0.3\linewidth}
        \includegraphics[width=\linewidth]{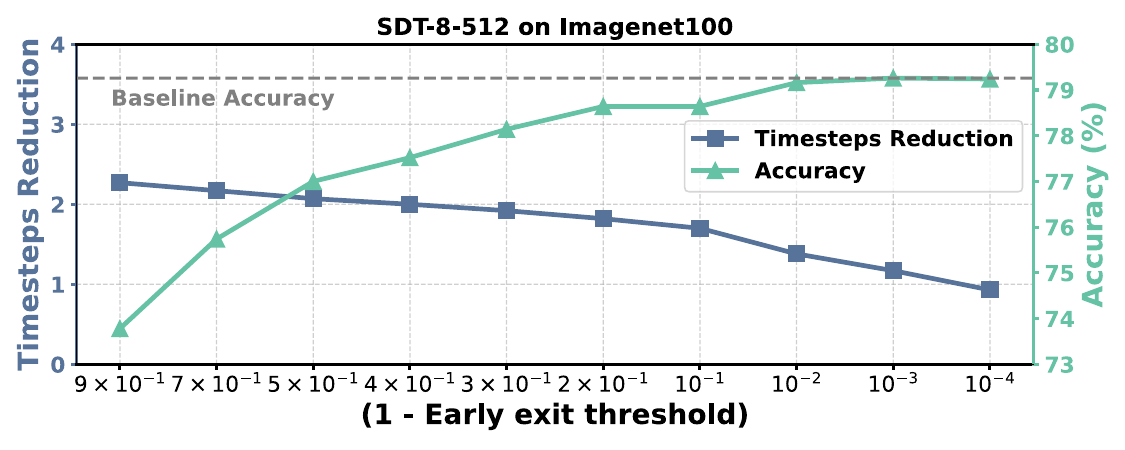}
    \end{minipage}
    \hfill
    \begin{minipage}{0.3\linewidth}
        \includegraphics[width=\linewidth]{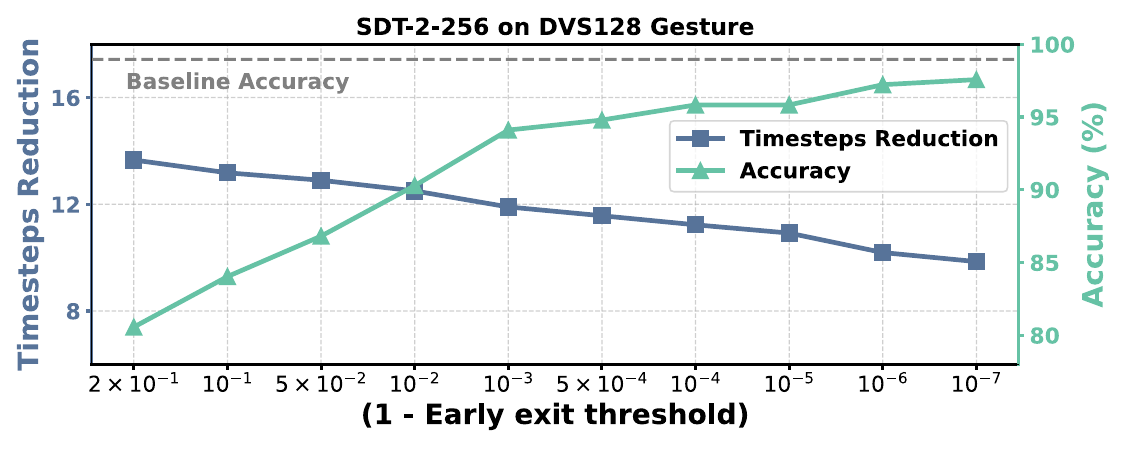}
    \end{minipage}
    \caption{Tradeoff of Early-Exit timestep reduction vs Accuracy across different datasets.}
    \label{fig:timestep_accuracy_combined}
\vspace{-3mm}
\end{figure*}

To assess temporal redundancy, we analyzed efficiency–accuracy trade-offs under timestep reduction (Figs.~\ref{fig:timestep_energy_combined}, \ref{fig:timestep_accuracy_combined}). Early-exit pruning consistently improved efficiency, yielding substantial reductions with $\sim$1\% accuracy loss: 45.6\% on CIFAR10-DVS, 23.3\% on ImageNet100, and 61.6\% on DVS-Gesture. Simpler datasets benefited from aggressive pruning, while ImageNet100 showed smaller but meaningful savings. This adaptive temporal processing lowers average timesteps per sample, reducing sequential MACs and energy.


\FloatBarrier

\begin{algorithm}[H]
\caption{Bayesian Optimization for Threshold Selection}
\label{alg:bayesianalgo}
\begin{flushleft}
\begin{framed}
\setcounter{equation}{0}

\textbf{Notation:}  
\begin{itemize}
  \item $\tau$: TAFT; $\beta$: CBET
  \item $\Theta = [\tau_{\min}, \tau_{\max}] \times [\beta_{\min}, \beta_{\max}]$: search space  
  \item $\theta = (\tau, \beta) \in \Theta$: candidate threshold configuration.  
  \item $g(\theta)$: scalarized objective combining accuracy and efficiency.  
  \item $\mathcal{D}_n = \{(\theta_i, y_i)\}_{i=1}^n$: dataset after $n$ evaluations.  
  \item $k(\cdot,\cdot)$: covariance kernel function of Gaussian Process (GP) surrogate model.   
  \item $m_{n-1}(\theta), s_{n-1}^2(\theta)$: GP posterior mean and variance at iteration $n-1$.  
  \item $g^+ = \max_{i=1,\ldots,n-1} y_i$: best observed value so far.  
  \item $\Phi(\cdot), \phi(\cdot)$: standard normal CDF and PDF.  
  \item $\text{EI}(\theta)$: Expected Improvement acquisition function.  
\end{itemize}

\medskip
\textbf{Initialize:} $\mathcal{D}_0 = \emptyset$, budget $N$, search space $\Theta$.  
Generate initial points via Latin Hypercube + corners: $\{\theta_1, \ldots, \theta_{n_0}\}$.

\medskip
\textbf{for } $i = 1, \ldots, n_0$ \textbf{ do}  
\[
y_i = g(\theta_i), \quad 
\mathcal{D}_i = \mathcal{D}_{i-1} \cup \{(\theta_i, y_i)\}
\]

\textbf{for } $n = n_0+1, \ldots, N$ \textbf{ do}  
\[
\begin{aligned}
m_{n-1}(\theta) &= k(\theta, \Theta)[K + \sigma^2 I]^{-1}y \\
s_{n-1}^2(\theta) &= k(\theta,\theta) - k(\theta,\Theta)[K + \sigma^2 I]^{-1}k(\Theta,\theta) \\
g^+ &= \max_{i=1,\ldots,n-1} y_i \\
z(\theta) &= \frac{m_{n-1}(\theta) - g^+}{s_{n-1}(\theta)} \\
\text{EI}(\theta) &= (m_{n-1}(\theta) - g^+)\Phi(z(\theta)) + s_{n-1}(\theta)\phi(z(\theta)) \\
\theta_n &= \arg\max_{\theta \in \Theta} \text{EI}(\theta) \\
y_n &= g(\theta_n), \quad 
\mathcal{D}_n = \mathcal{D}_{n-1} \cup \{(\theta_n, y_n)\}
\end{aligned}
\]

\textbf{end for}  

\medskip
Extract Pareto frontier $\mathcal{P}$ via non-dominance filtering of $\mathcal{D}_N$.

\end{framed}
\end{flushleft}
\end{algorithm}

\FloatBarrier

\begin{figure}
    \centering
    \includegraphics[width=\linewidth]{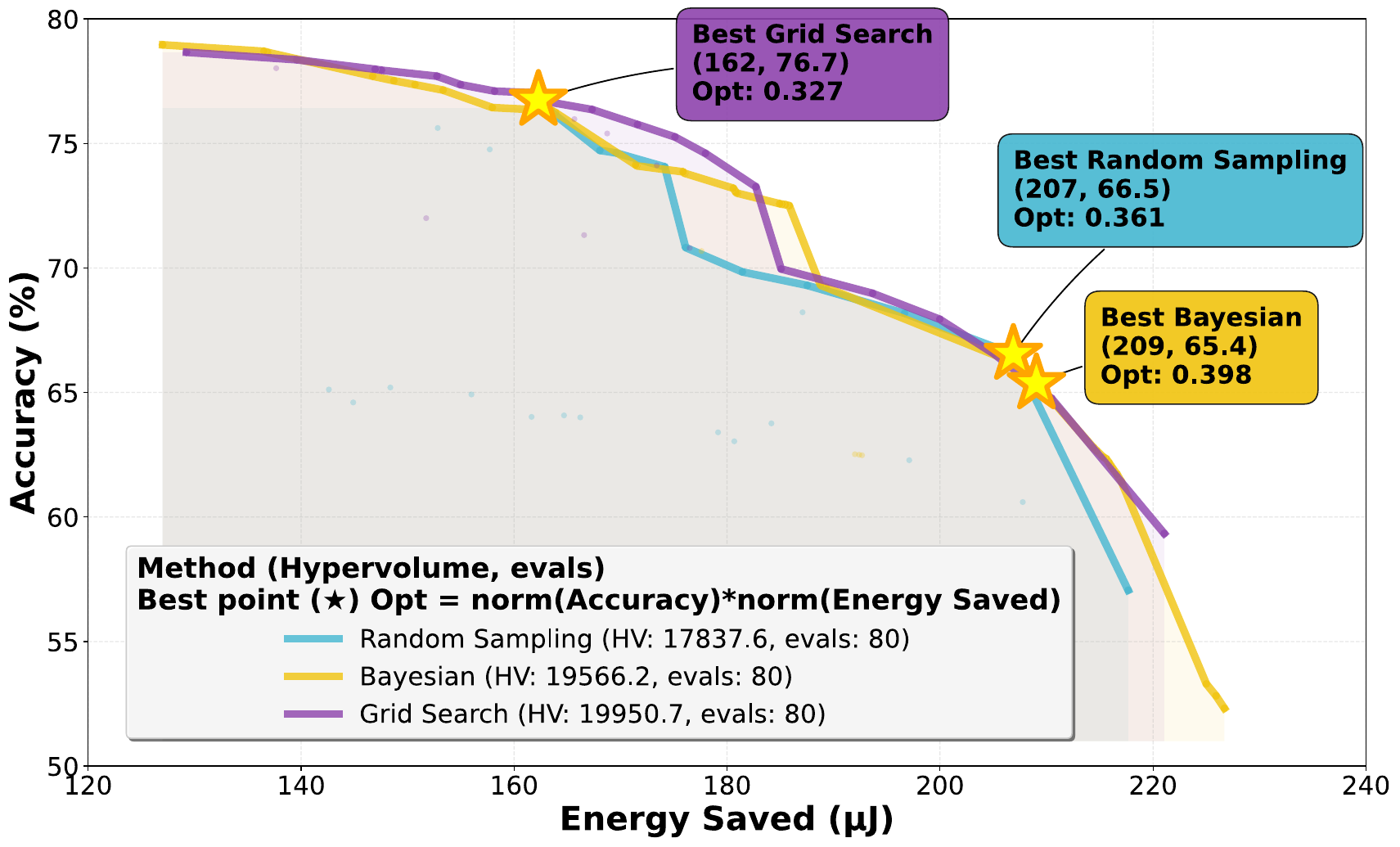}
    \caption{Pareto frontiers comparison of different threshold optimization methods on SDT-8-512.}
    \label{fig:pareto_frontiers_comparison}
\end{figure}

Threshold selection optimization with SDT-8-512 on ImageNet100 (80-evaluation budget, $9.577~\mu$J per skipped layer, $76.135~\mu$J per saved timestep, min accuracy 47\% across four timesteps) demonstrated complementary strategies. Grid search achieved the best hypervolume (19.95k) but required exhaustive trials, while Bayesian optimization rapidly identified the best configuration (metric 0.398). Both outperformed random and manual baselines (Fig.~\ref{fig:pareto_frontiers_comparison}), underscoring the value of principled threshold optimization.

    \section{Conclusion}
In this work, we presented ASTER, a hybrid analog–digital Processing-in-Memory (PIM) accelerator designed to efficiently execute spiking transformers for event-driven vision tasks. Our architecture is carefully co-optimized with the sparse, binary nature of spiking data to reduce memory access overhead and maximize data reuse. The resulting design supports direct integration with event-based encoders and outputs compatible with symbolic reasoning backends, forming a complete neurosymbolic pipeline. Comprehensive evaluations on representative visual reasoning tasks demonstrate up to 467× energy reduction compared to edge GPUs (Jetson Orin Nano) and 1.86× savings over prior PIM-based spiking transformer accelerators, all while maintaining competitive accuracy. These results highlight the viability of ASTER as a low-power, high-efficiency accelerator for real-time event-driven applications on edge devices.

\clearpage
\bibliographystyle{ieeetr}
\bibliography{mybibliography}
\end{document}